\documentclass[a4paper,11pt]{article}
\pdfoutput=1 

\usepackage{jheppub}
\usepackage[T1]{fontenc} 

\usepackage[dvipsnames]{xcolor}

\usepackage{enumitem}
\usepackage{mathrsfs}

\newcommand{\bei}{\begin{itemize}}
\newcommand{\eei}{\end{itemize}}
\newcommand{\bee}{\begin{enumerate}}
\newcommand{\eee}{\end{enumerate}}
\newcommand{\beeL}{\begin{enumerate}[label=(\Alph*)]}
\newcommand{\beel}{\begin{enumerate}[label=(\alph*)]}
\newcommand{\beeR}{\begin{enumerate}[label=(\Roman*)]}
\newcommand{\beer}{\begin{enumerate}[label=(\roman*)]}
\newcommand{\beeLd}{\begin{enumerate}[label=\Alph*.]}
\newcommand{\beeld}{\begin{enumerate}[label=\alph*.]}
\newcommand{\beeRd}{\begin{enumerate}[label=\Roman*.]}
\newcommand{\beerd}{\begin{enumerate}[label=\roman*.]}

\newcommand{\bal}{\begin{equation}\begin{aligned}}
\newcommand{\eal}{\end{aligned}\end{equation}}

\newcommand{\ov}{\over}
\newcommand{\g}{\gamma}

\def\sgn{{\text{sgn}}}

\definecolor{grey}{rgb}{0.4,0.4,0.5}
\definecolor{darkgreen}{rgb}{0,0.5,0}
\definecolor{darkred}{rgb}{0.6,0.0,0}
\definecolor{lightbrown}{rgb}{1,0.9,0.8}
\definecolor{brown}{rgb}{0.6,0.3,0.3}
\definecolor{darkblue}{rgb}{0,0,0.5}
\definecolor{darkmagenta}{rgb}{0.5,0,0.5}

\def\bZ{{\mathbb Z}}

\newcommand{\la}{\label}
\def\a {\alpha}
\def\b {\beta}

\newcommand{\B}{{\scriptscriptstyle\text{B}}}
\newcommand{\F}{{\scriptscriptstyle\text{F}}}
\renewcommand{\L}{{\scriptscriptstyle\text{L}}}
\newcommand{\R}{{\scriptscriptstyle\text{R}}}

\def\mC{{\mathscr C}}

\def\cK{{\cal K}}

\def\cR{{\cal R}}

\def\ket{\rangle}

\def\w{{\omega}}
\def\bS{{\mathbb S}}
\newcommand{\A}{{\scriptscriptstyle\text{A}}}
\newcommand{\C}{{\scriptscriptstyle\text{C}}}
\newcommand{\D}{{\scriptscriptstyle\text{D}}}
\def\bI{{\mathbb I}}
\newcommand{\bA}{\mathbf{A}}
\newcommand{\G}{{\scriptscriptstyle\text{G}}}

\title{\boldmath Exchange relations and crossing}

\author[1,4]{Sergey Frolov,}
\author[2]{Davide Polvara,}
\author[3,4]{Alessandro Sfondrini}

\affiliation[1]{Hamilton Mathematics Institute and School of Mathematics Trinity College, Dublin 2, Ireland.}
\affiliation[2]{II. Institut f\"ur Theoretische Physik,  Universit\"at Hamburg, Luruper Chaussee 149, 22761 Hamburg, Germany.}
\affiliation[3]{Dipartimento di Fisica e Astronomia, Universit\`a degli Studi di Padova, via Marzolo 8,
35131 Padova, Italy.}
\affiliation[4]{Istituto Nazionale di Fisica Nucleare, Sezione di Padova, via Marzolo 8, 35131 Padova,
Italy.}

\emailAdd{frolovs@maths.tcd.ie}
\emailAdd{davide.polvara@gmail.com}
\emailAdd{alessandro.sfondrini@unipd.it}

\abstract{We discuss the scattering matrix of two-dimensional integrable QFTs whose fields obey non-trivial exchange relations. We show that  crossing equations for such models have to be modified, and propose their consistent modification. This modification opens the way to constructing new integrable S~matrices.
As a check, we  consider the crossing equations for the $SU(N)$ chiral Gross-Neveu model, and for the $\Phi_{21}$ deformation of the tricritical Ising model, finding an agreement with the existing proposals. 
Finally, we reconsider the crossing equations for massless excitations of the mixed-flux $AdS_3\times S^3\times T^4$ light-cone gauge superstring sigma model, and conjecture that the massless excitations satisfy non-trivial exchange relations. This changes the crossing equations and leads to a simpler massless dressing factor.
}

\begin{document} \begin{flushright}\small{ZMP-HH/25-9}\end{flushright}
\maketitle
\section{Introduction}
\label{sec:introduction}
It is well-known that the fields of a quantum field theory in two spacetime dimension do not have to be either bosons or fermions. In general they may satisfy non-trivial exchange relations \cite{Leinaas:1977fm}. This has profound consequence for scattering: in particular, the S~matrix (or more precisely, the \textit{physical S~matrix}~$\mathbb{S}$) will not be analytical; this has been discussed in detail especially in the case of of integrable QFTs, where (factorised) scattering can be studied in full detail~\cite{Swieca:1976dv,Koberle:1979ne,Karowski:1980kq,Smirnov:1990pr}.
In this paper we consider two-dimensional integrable QFTs --- relativistic and non-relativistic --- and show that if the fields obey non-trivial exchange relations then in general the crossing equations have to be modified, and propose their consistent modification. One can also reverse the logic and claim that as soon as the crossing equations of a model are non-standard the exchange relations are non-trivial. 

 The simplest example is the $SU(2)$ chiral Gross-Neveu (CGN) model~\cite{Gross:1974jv} where the proposed S-matrix satisfies the crossing equations ``with a wrong sign'', with respect to what one may na\"ively expect. The fields of the model obey non-trivial exchange relations~\cite{Koberle:1979ne}, and this is the reason for the change of the sign. Moreover, if one does not assume particular exchange relations from the very beginning then the minimal solution of our modified crossing equation also determines the exchange relations.  Similar considerations apply to the $SU(N)$ CGN model, where the na\"ive crossing equations are modified by a phase~\cite{Berg:1977dp,Berg:1978zn,Abdalla:1979sa,Koberle:1979ne}.
 
Another example is 
the $\Phi_{21}$ deformation of the tricritical Ising model originally discussed in \cite{Zamolodchikov:1990xc,Colomo:1991gw,Smirnov:1991uw}
where the S-matrix proposed in \cite{Colomo:1991gw,Smirnov:1991uw}
does not satisfy the standard crossing equations. This example is even more non-trivial than the CGN model, as the modification of crossing is not just an overall phase factor, but it has instead a non-trivial matrix structure.
It was argued
recently in~\cite{Copetti:2024rqj,Copetti:2024dcz} that the modification of the crossing equations is related to the invariance of the model under categorical symmetries. 
Here we show that the S-matrix proposed in \cite{Zamolodchikov:1990xc} implies non-trivial exchange relations, and our modified crossing equation and its minimal solution agree with the ones in~\cite{Colomo:1991gw,Smirnov:1991uw,Copetti:2024rqj,Copetti:2024dcz}. 

We also reconsider the crossing equations for massless excitations of the mixed-flux $AdS_3\times S^3\times T^4$ light-cone gauge superstring sigma model (see~\cite{Sfondrini:2014via,Demulder:2023bux}, for reviews). The crossing equations for the model were derived in \cite{Lloyd:2014bsa} under the assumption that the particles are either bosons or fermions. The solution to the crossing equations
in the Ramond-Ramond (RR) case \cite{Frolov:2021fmj} contains a function which is only necessary in this case. 
Apparently, the presence of that function is also incompatible with the proposed Quantum Spectral Curve (QSC) for the RR model~\cite{Ekhammar:2021pys,Cavaglia:2021eqr} as was recently argued in \cite{Ekhammar:2024kzp}. In this paper we conjecture that the massless excitations satisfy non-trivial exchange relations similar to the ones for the $SU(2)$ CGN model. This changes the sign of the crossing equations and leads to a simpler massless dressing factor compatible both with perturbative calculations and the QSC.

The plan of the paper is as follows. In section 2 we discuss the  most general exchange relations without an assumption of the relativistic invariance. We consider two-particle scattering and relate the physical S-matrix $\bS(p_1,p_2)$ to the S-matrix $S(p_1,p_2)$ which appears in the Zamolodchikov-Faddeev (ZF) algebra  \cite{Zamolodchikov:1978xm, Faddeev:1980zy}. In the limit of trivial scattering, e.g. $\theta_1-\theta_2\to\infty$ for relativistic models, the physical S-matrix becomes the identity while the ZF one $S(p_1,p_2)$ becomes the $\cR$-matrix which appears in the exchange relations. We assume the existence of a singlet state which scatters ``trivially'' with the ZF operators, and use the ZF algebra to derive the modified crossing equations. At the end of section 2 we discuss the generalisation of the formalism to the case of a restricted Hilbert space of states where some of the states are unphysical.  In section 3 we consider the $SU(N)$ CGN model, and the $\Phi_{21}$ deformation of the tricritical Ising model, and derive their modified crossing equations. The section 4 is devoted to the revision of the crossing equations for massless exitations of the mixed-flux  $AdS_3\times S^3\times T^4$ light-cone gauge superstring sigma model.
We conclude in section~5.

\section{Non-trivial exchange relations and the S matrix}
\label{sec:Smatrix}

In this section we review some properties of 2d models for which scattering states are created by  operators satisfying non-trivial exchange relations. In order to apply the formalism to the models appearing in the AdS/CFT context we do not assume  relativistic invariance.

\subsection{Exchange relations}

Let us consider a system of $K$ particles with 
 {\it in} and {\it out} creation and annihilation operators satisfying the following exchange relations \cite{Koberle:1979ne,Karowski:1980kq,Smirnov:1990pr}\footnote{We follow \cite{Smirnov:1990pr} and use the same exchange relations for  {\it in} and {\it out} operators. It allows one to define the scattering matrix in the usual way.}
\bal\la{algcR}
{\bf a}_1^{\dagger} {\bf a}_2^{\dagger}&=  {\bf a}_2^{\dagger} {\bf a}_1^{\dagger}\, \cR_{12}\,,\quad {\bf a}_1{\bf a}_2=  \cR_{12}\, {\bf a}_2 {\bf a}_1 \,,\quad
{\bf a}_1 {\bf a}_2^{\dagger}=  {\bf a}_2^{\dagger} \,\cR_{21}\, {\bf a}_1 + \delta_{12}\,.
\eal
Here we use the notations from  review \cite{Arutyunov:2009ga}, section 3
\bal
\cR_{12} = \cR_{\A\B}^{\C\D}(p_1,p_2)E_\C{}^\A\otimes E_\D{}^\B\,,\ \ \cR_{21} = \cR_{\A\B}^{\C\D}(p_2,p_1)E_\D{}^\B \otimes E_\C{}^\A\,,\ \ \delta_{12}= \delta(p_1-p_2) E_\A\otimes E^\A
\eal
where $E_\C{}^\A=E_\C\otimes E^\A$ are matrix unities, and $E_\C$ and $E^\A$ are canonical basis columns  and rows, respectively. Then,
\bal
{\bf a}^{\dagger} = a_\A^{\dagger}  E^\A\,,\quad {\bf a} = a^\A E_\A\,,
\eal
are a row  and a column  of creation and annihilation operators, and
\bal
{\bf a}_1^{\dagger} {\bf a}_2^{\dagger} = a_\A^{\dagger}(p_1)a_\B^{\dagger}(p_2)  E^\A\otimes E^\B\,,\\
{\bf a}_2^{\dagger} {\bf a}_1^{\dagger} = a_\B^{\dagger}(p_2)a_\A^{\dagger}(p_1)  E^\A\otimes E^\B\,, 
\eal
and similar formulae for other expressions. Above and in what follows if a formula is valid for both the  {\it in} and  {\it out}  operators we just use $a_\A^{\dagger}$, $a^{\A}$.

In components the exchange relations \eqref{algcR} take the form
\bal\la{algcR2}
a_\A^{\dagger}(p_1) a_\B^{\dagger}(p_2)&=  a_\D^{\dagger}(p_2) a_\C^{\dagger}(p_1)\cR_{\A\B}^{\C\D}(p_1,p_2)\,,\\
a^\A(p_1) a^\B(p_2)&= \cR^{\A\B}_{\C\D}(p_1,p_2)  a^\D(p_2) a^\C(p_1)\,,
\\
a^\A(p_1) a_\B^{\dagger}(p_2)&=  a_\D^{\dagger}(p_2)\cR_{\B\C}^{\D\A}(p_2,p_1) a^\C(p_1)+\delta_\B^\A \delta(p_1-p_2)\,.
\eal
Equations \eqref{algcR}, \eqref{algcR2} define the same algebra as the Zamolodchikov-Faddeev one \cite{Zamolodchikov:1978xm, Faddeev:1980zy}, and $\cR_{12}$ satisfies the usual requirements of physical unitarity, braiding unitarity and the Yang-Baxter equation
\bal
\cR_{12}^\dagger = \cR_{12}^{-1}= \cR_{21} \,,\quad \cR_{12}\cR_{13}\cR_{23}=\cR_{23}\cR_{13}\cR_{12}\,.
\eal
We also impose the following condition
 \bal\la{algcR3}
 \sum_{\F,\G} \w_\F(p_1)\cR_{\F\G}^{\C\D}(p_1,p_2)\cR_{\B\A}^{\G\F}(p_2,p_1) =  \w_\A(p_1)\delta_\A^\C\delta_\B^\D\,,
 \eal
where $\w_\A(p)$ is the energy (dispersion relation) of the $A$-th particle at asymptotic infinity. This condition follows from the braiding unitarity and the fact that for particles with different dispersion relations only transmission amplitudes do not vanish. 

If a model is relativistic invariant then  $\cR_{12}$ must depend on the difference of rapidities 
$\theta_{12}\equiv \theta_1-\theta_2$, and 
it is usually chosen to be piecewise constant
\bal\la{relcR12}
\cR_{12} = R_{12}\,u(\theta_{12})+ R_{21}^{-1}\,u(\theta_{21})\,,
\eal
where $u(\theta)$ is the Heaviside (unit-step) function, while $R_{12}$ and $R_{21}$ are constant unitary matrices 
  satisfying the YB equation.  In general, they do not satisfy the braiding unitarity: $R_{21}^{-1}\neq R_{12}$.
  The simplest and the most studied case corresponds to a diagonal $\cR_{12}$
\bal\la{diagR12}
\cR_{\A\B}^{\C\D}(p_1,p_2)= \delta_{\A}^{\C}\delta_{\B}^{\D}\,e^{-2\pi i \,s_{\text{$\A\B$}}\,\sgn(\theta_{12}) }\,,
\eal
where $s_{\A\B}=s_{\B\A}$  is a  real number.   We refer to it as to the relative spin of the operators $a _\A^\dagger$, $a^\A$ and $a _\B^\dagger$, $a^\B$. The relative spin is obviously defined modulo~1.
Remark that in non-relativistic case we can still use \eqref{relcR12} with $\theta_{12}$ replaced by the relative group velocity, $v_{12}=v_1-v_2$, where $v=\partial \omega/\partial p$.

As usual, the Hamiltonian of the system in terms of the {\it in} and {\it out} creation and annihilation operators is given by
\bal
H = \int\, dp\, \w_\A(p)\,  a^{\dagger}_\A(p) a^\A(p) \,.
\eal
By using the exchange relations  \eqref{algcR2}, \eqref{algcR3}
 one can easily check that the eigenvectors are 
\bal
  a^{\dagger}_{\A_1}(p_1)  a^{\dagger }_{\A_2}(p_2)\cdots   a^{\dagger}_{\A_n}(p_n)|0\ket\,,
\eal
with the eigenvalues $\sum_i \w_{\A_i}(p_i)$. 

\subsection{Scattering matrix}

We follow the standard route and  define the {\it in} and {\it out} states 
\bal
|p_1,p_2,\ldots, p_n\ket_{\A_1\A_2\cdots \A_n}^{\rm in} &=  a^{\dagger {\rm in}}_{\A_1}(p_1)  a^{\dagger {\rm in}}_{\A_2}(p_2)\cdots   a^{\dagger {\rm in}}_{\A_n}(p_n)|0\ket\,,
\\
|p_1,p_2,\ldots, p_n\ket_{\A_1\A_2\cdots \A_n}^{\rm out} &=  a^{\dagger {\rm out}}_{\A_1}(p_1)  a^{\dagger {\rm out}}_{\A_2}(p_2)\cdots   a^{\dagger {\rm out}}_{\A_n}(p_n)|0\ket\,,
\eal
and the S-matrix operator 
\bal
|p_1,p_2,\ldots, p_n\ket_{\A_1\A_2\cdots \A_n}^{\rm in} &= \bS\cdot
|p_1,p_2,\ldots, p_n\ket_{\A_1\A_2\cdots \A_n}^{\rm out} \,.
\eal
Clearly, we can expand initial states over a basis of final states and vice versa. In particular, for the two-particle {\it in} and {\it out} states we get either
\bal
|p_1,p_2\ket_{\A\B}^{\rm in} &= \bS\cdot
|p_1,p_2\ket_{\A\B}^{\rm out} =\bS_{\A\B}^{\C\D}(p_1,p_2)\,|p_1,p_2\ket_{\C\D}^{\rm out} \,,
\eal
or equivalently, 
\bal
\bS\cdot
|p_1,p_2\ket_{\A\B}^{\rm in} =\bS_{\A\B}^{\C\D}(p_1,p_2)\,|p_1,p_2\ket_{\C\D}^{\rm in}\,.
\eal
Here we assume that the set of momenta of the two scattering particles does not change in the scattering process, and we also order the particle's velocities, $v_k=\partial\w_{A_k}(p_k)/\partial p_k$, in decreasing order  $v_1>v_2>\cdots>v_n$ to take into account the particle's statistics. $\bS_{\A\B}^{\C\D}(p_1,p_2)$ are the matrix elements of the physical two-particle S-matrix.  Clearly, the S-matrix commutes with the Hamiltonian, and in the absence of interaction $\bS =\bI$, and $\bS_{\A\B}^{\C\D}=\delta_{\A}^{\C}\delta_{\B}^{\D}$. 

To describe the scattering process, we introduce the ZF operators and the {\it in}-basis and the {\it out}-basis of asymptotic states as
\bal
|p_1,p_2,\ldots, p_n\ket_{\A_1\A_2\cdots \A_n}^{\rm in} &=  A^{\dagger }_{\A_1}(p_1) \cdots   A^{\dagger}_{\A_n}(p_n)|0\ket\,,\qquad v_1>v_2>\cdots>v_n\,,
\\
|p_1,p_2,\ldots, p_n\ket_{\A_1\A_2\cdots \A_n}^{\rm out} & = \prod_{i<j} \cR_{\A_i\A_j}^{\B_i\B_j}(p_i,p_j)\, A^{\dagger }_{\B_n}(p_n) \cdots   A^{\dagger}_{\B_1}(p_1)|0\ket\,,\qquad v_1>v_2>\cdots>v_n\,.
\eal
These bases are obtained by replacing $ a^{\dagger {\rm in}}_{\A_k}(p_k)$ with $ A^{\dagger }_{\A_k}(p_k)$ in the {\it in} state, and first reordering the $ a^{\dagger {\rm out}}_{\A_k}(p_k)$ in the {\it out} state and then replacing them with $ A^{\dagger}_{\A_k}(p_k)$.

In terms of the ZF operators one finds 
\bal
 A^{\dagger }_{\A}(p_1)  A^{\dagger}_{\B}(p_2)|0\ket =\cR^{\C\D}_{\F\G}(p_1,p_2) \,\bS_{\A\B}^{\F\G}(p_1,p_2)\, A^{\dagger }_{\D}(p_2)    A^{\dagger}_{\C}(p_1)|0\ket\,,
\eal
and dropping $|0\ket$, we get the ZF algebra
\bal
 A^{\dagger }_{\A}(p_1)  A^{\dagger}_{\B}(p_2) =A^{\dagger }_{\D}(p_2)    A^{\dagger}_{\C}(p_1)\, S_{\A\B}^{\C\D}(p_1,p_2)\,,
\eal
where
\bal\la{defS12}
S_{\A\B}^{\C\D}(p_1,p_2)\equiv \cR^{\C\D}_{\F\G}(p_1,p_2) \,\bS_{\A\B}^{\F\G}(p_1,p_2) \,,\qquad S_{12}=\cR_{12}\,\bS_{12}  \,,\quad v_1>v_2\,.
\eal
Let us stress that it is $S_{\A\B}^{\C\D}(p_1,p_2)$ that satisfies the usual requirements for  S-matrix elements, in particular some kind of analyticity,  the braiding unitarity, YB equation and {\it modified} crossing symmetry conditions which we will discuss at the end of this section. We will refer to $S_{\A\B}^{\C\D}(p_1,p_2)$ as the ZF S~matrix, to distinguish it from the physical S~matrix~$\mathbb{S}_{\A\B}^{\C\D}(p_1,p_2)$. Taking into account that in the absence of interaction $\bS_{12}\to\bI_{12}$, we get

\bal
S_{12}\xrightarrow{\text{trivial scattering}} \cR_{12}\,,\quad v_1>v_2\,.
\eal
Extending the relation \eqref{defS12} to all values of $p_1,p_2$, we get
\bal
 \bS_{12}= \cR_{21}\,S_{12}\,,
\eal
which shows that $\bS_{12}$ is in general non-analytic.  In particular, in the simplest case \eqref{diagR12} of the diagonal $\cR_{12}$, the physical S-matrix is given by
\bal
 \bS_{\A\B}^{\C\D}(p_1,p_2)= S_{\A\B}^{\C\D}(p_1,p_2)\, e^{+2\pi i s_{\text{$\C\D$}} \sgn(v_1-v_2)} \,,
\eal
which is non-analytic unless $s_{\A\B}=0$ mod$\, \tfrac{1}{2}$.
Note also that in the absence of interaction 
\bal
S_{\A\B}^{\C\D}(p_1,p_2)\to \delta_{\A}^{\C}\delta_{\B}^{\D}\,e^{-2\pi i s_{\text{$\C\D$}} \sgn(v_1-v_2)}\,.
\eal

\subsection{Symmetries}

Let us now assume that the $K$ particles are split into two groups\footnote{Generalisation to any number of groups is straightforward.} so that  particles in a group have the same dispersion relation 
\bal
\w_{k}(p) = \w^{(1)}(p)\,,\quad k=1,2,\ldots, N\,,\qquad \w_{\a}(p) = \w^{(2)}(p)\,,\quad \a=N+1,N+2,\ldots, K\,.
\eal
Then, the Hamiltonians $H$ takes the form
\bal
H = H^{(1)}+H^{(2)}\,,\quad H^{(1)}= \int dp\,  \w^{(1)}(p) a^{\dagger}_{i}(p)a^i(p)\,,\quad H^{(2)}= \int dp\,  \w^{(2)}(p) a^{\dagger}_{\a}(p)a^\a(p)\,,
\eal 
and the $\cR$-matrix (and S-matrix) has the following non-vanishing elements
\bal
\cR_{ij}^{kl}\,,\quad \cR_{i\b}^{k\rho}\,,\quad \cR_{\a j}^{\g l}\,,\quad \cR_{\a\b}^{\g\rho}\,. 
\eal
The braiding unitarity then implies
\bal
\cR_{nm}^{kl}(p_1,p_2)\cR_{ji}^{mn}(p_2,p_1) &= \delta^k_i\delta^l_j\,,\quad \cR_{\nu\mu}^{\g\rho}(p_1,p_2)\cR_{\b\a}^{\mu\nu}(p_2,p_1) =\delta^\g_\a \delta^\rho_\b\,,\quad 
\\
\cR_{n\mu}^{k\rho}(p_1,p_2)\cR_{\b i}^{\mu n}(p_2,p_1) &=\delta^k_i \delta^\rho_\b\,,\quad
 \cR_{\nu m}^{\g l}(p_1,p_2)\cR_{j\a}^{m\nu}(p_2,p_1) = \delta^\g_\a\delta^l_j\,.\quad 
\eal
By using the relations, we find the commutators of $H^{(1)}$ and $H^{(2)}$ with the oscillators
\bal
\relax
[ \, H^{(1)} \,,\, a_i^\dagger(p)\, ]&=\w^{(1)}(p)\, a_i^\dagger(p)\,,\quad [ \, H^{(1)} \,,\, a_\a^\dagger(p)\, ]=0\,,
\\
[ \, H^{(2)} \,,\, a_\a^\dagger(p)\, ]&=\w^{(2)}(p)\, a_\a^\dagger(p)\,,\quad [ \, H^{(2)} \,,\, a_i^\dagger(p)\, ]=0\,.
\eal
These relations imply that the Hamiltonian $H$
 commutes with the following charges
\bal
L_i{}^j &= \int dp\, a^{\dagger}_{i}(p)a^j(p)\,,\qquad  i,j=1,\ldots, N \,,
\\
L_\a{}^\b &= \int dp\,  a^{\dagger}_{\a}(p) a^\b(p)\,,\qquad  \a,\b=N+1,\ldots, K\,.
\eal 
The charges $L_i{}^j$ and $L_\a{}^\b$ above are defined in terms of either \textit{in} or \textit{out} oscillators; if they form a Lie (super-)algebra and the oscillators fit in its representation, we can unambiguously determine their action on the ZF operators $A^\dagger_i, A^j$. However, in general this is not the case,%
\footnote{$L_i{}^j$ and $L_\a{}^\b$ may not even (anti-)commute with each other.}
in which case it is harder to determine the constraints they impose on the ZF S~matrix. 
Still, their action on multi-particle states \textit{in} (respectively, \textit{out}) states can be easily found, e.g.
\bal
 L_i{}^j  a^{\dagger}_{\A}(p) a^{\dagger}_{\B}(p')|0\ket &= \delta_\A^j  a^{\dagger}_{i}(p) a^{\dagger}_{\B}(p')|0\ket + \cR_{\A\B}^{\D j}(p,p')\cR_{i\D}^{\C' \C}(p',p)  a^{\dagger}_{\C}(p) a^{\dagger}_{\C'}(p')|0\ket\,,
 \\
  L_\a{}^\b  a^{\dagger}_{\A}(p) a^{\dagger}_{\B}(p')|0\ket &= \delta_\A^\b  a^{\dagger}_{\a}(p) a^{\dagger}_{\B}(p')|0\ket + \cR_{\A\B}^{\D \b}(p,p')\cR_{\a\D}^{\C' \C}(p',p)  a^{\dagger}_{\C}(p) a^{\dagger}_{\C'}(p')|0\ket\,.
\eal
Since $L_i{}^j$ act closely in the space of two-particle states, one can derive nontrivial constraints on the form of $\cR_{\A\B}^{\C\D}$, see e.g. review \cite{Arutyunov:2009ga} where the derivation was done for the scattering matrix of (anti)-commuting fields. If $L_i{}^j$ generate a (super-)Lie algebra then 
 the S-matrix elements $S_{\A\B}^{\C\D}$ satisfy \textit{the same} constraints. 

Let us now discuss in more detail the simplest case of a diagonal $\cR_{12}$,
\bal\la{diagR12b}
\cR_{\A\B}^{\C\D}(p_1,p_2)= \delta_{\A}^{\C}\delta_{\B}^{\D}\,e^{-2\pi i \,s_{\text{$\A\B$}}\,\sgn(v_{12}) }\,.
\eal
In this case it is easy to show that
\bal
 L_\A{}^\B   L_\C{}^\D &= \int dpdp'\,  a^{\dagger}_{\A}(p) a^\B(p) \,  a^{\dagger}_{\C}(p') a^\D(p')=\delta_\C{}^\B  L_\A{}^\D - \delta_\A{}^\D  L_\C{}^\B 
 \\
& +
 \int dpdp'\,  e^{+2\pi i (s_{\A \D}-s_{\A \C}-s_{\B \D}+s_{\B \C}) \sgn(v-v')} a^{\dagger}_{\C}(p') a^\D(p') a^{\dagger}_{\A}(p) a^\B(p) \,.
\eal
Again, in general, the operators $L_\A{}^\B$ and $L_\C{}^\D$ do not satisfy any (super-)Lie algebra relations.
Nevertheless, we see that if for some indices A, B, C, D 
\bal\la{spincond}
s_{\A \D}-s_{\A \C}-s_{\B \D}+s_{\B \C}=0\quad \text{mod}\, 1\,,
\eal
then $L_\A{}^\B$ and   $L_\C{}^\D$ commute while if 
\bal\la{spincond2}
s_{\A \D}-s_{\A \C}-s_{\B \D}+s_{\B \C}={1\ov2}\quad \text{mod}\, 1\,,
\eal
they anti-commute.
Thus, if 
\bal
s_{il}-s_{ik}-s_{jl}+s_{jk}=
\left\{
\begin{array}{ccc}
 0\quad \text{mod}\, 1\,, &   \text{for some} & 
 i,j,k,l   \\
   {1\ov2}\quad \text{mod}\, 1\,, &   \text{for others} & 
 i,j,k,l   
\end{array}
\right.
\,,
\eal
then the charges $L_i{}^j $  satisfy a $u(M|N-M)$ super-algebra relations
\bal
\left[L_i{}^j \,,\, L_k{}^l\, \right]_\pm =\delta_k{}^j  L_i{}^l - \delta_i{}^l   L_k{}^j \,.
\eal
The action of $L_i{}^j$ on  two-particle states is given  by
\bal\la{coprod}
 L_i{}^j  a^{\dagger}_{\A}(p) a^{\dagger}_{\B}(p')|0\ket = \delta_\A^j  a^{\dagger}_{i}(p) a^{\dagger}_{\B}(p')|0\ket + \delta_\B^j e^{+2\pi i( s_{i\A}-s_{j\A})\sgn(v-v')} a^{\dagger}_{\A}(p) a^{\dagger}_{i}(p')|0\ket\,,
\eal
and, therefore, in general  we get a non-standard co-product.

Let us see what we get in the $\mathfrak{u}(2)$ case. It is easy to check that \eqref{spincond} implies
\bal
s_{12}=s_{21}={s_{11}+s_{22} + n\ov2}\,,\quad n\in\bZ\,.
\eal
Then, \eqref{coprod} gives
\bal\la{coprod11}
 L_i{}^i  a^{\dagger}_{k}(p) a^{\dagger}_{l}(p')|0\ket = \delta_k^j  a^{\dagger}_{i}(p) a^{\dagger}_{l}(p')|0\ket + \delta_l^j  a^{\dagger}_{k}(p) a^{\dagger}_{i}(p')|0\ket\,.
\eal
\bal\la{coprod12}
& L_1{}^2  a^{\dagger}_{k}(p) a^{\dagger}_{l}(p')|0\ket = \delta_k^j  a^{\dagger}_{i}(p) a^{\dagger}_{l}(p')|0\ket + \delta_l^j e^{+2\pi i( s_{1k}-s_{2k})\sgn(v-v')}  a^{\dagger}_{k}(p) a^{\dagger}_{i}(p')|0\ket\,,
\eal
\bal
&s_{11}-s_{21}=s_{11}-s_{12}={ s_{11} - s_{22} - n\ov2}\,,\quad s_{12}-s_{22}={ s_{11} - s_{22} + n\ov2}=s_{11}-s_{21} +n\,.
\eal
Thus, if we want to have the standard plus sign between the two terms in \eqref{coprod12} then 
we can simply set all mutual spins to be equal to each other: $s_{11} = s_{22}=s_{12}$. On the other hand to get
the minus sign  in \eqref{coprod12}  we  set
\bal
{ s_{11} - s_{22} - n\ov2}={ 1\ov2}\,,\quad n=s_{11} - s_{22} -1\,,\quad s_{12}=s_{11}-{1\ov2}\,,\quad s_{11} - s_{22}\in\bZ\,,
\eal
and simple solutions are 
\bal
s_{11}=s_{22}=0\,,\  s_{12}=-{1\ov2}\,,\quad \text{or} \quad s_{11}=s_{22}={1\ov2}\,,\  s_{12}=0\,,\quad \text{or} \quad s_{11}=s_{22}={1\ov4}\,,\  s_{12}=-{1\ov4}\,.
\eal
 The minus sign in \eqref{coprod12} means that
we deal with the $\mathfrak{U}_{-1}(\mathfrak{su}(2))$ Hopf algebra, that was the choice made in  \cite{Lukyanov:1993pn} for the $SU(2)$ Thirring model.

\subsection{Modified crossing equations}

We define the crossing transformation as an analytic continuation of momentum $p$ to $\bar p\equiv -p$ such that the dispersion relation $\w_\A(p)$ changes its sign too: $\w_\A(p)\to -\w_\A(p)$. Thus, it replaces a particle with its anti-particle, and the crossing symmetry relates the scattering matrix of particles to the one of a particle with an anti-particle. 

A model with non-trivial exchange relations in general satisfies modified crossing equations. To derive these equations we assume 
the existence of a singlet state in the ZF algebra, which is annihilated by all the symmetry charges, following an idea put forward in~\cite{Beisert:2005tm} in the context of $\mathcal{N}=4$ supersymmetric Yang-Mills integrability.  
Such a state, which we label by $\Xi$, is defined by combining particles and anti-particles with opposite energies and momenta as follows
\bal
\Xi(p)=\bA^\dagger(p)\mC \bA^{\dagger t}(\bar p)\,,\quad \bar p= -p\,,
\eal
where $\bA^\dagger(p)=A^\dagger_\A(p)\,E^\A$, $\bA^{\dagger t}(p)=A^\dagger_\A(p)\,E_\A$ and $\mC =(\mC^{\A}{}_{\B})$ is found from the condition that $\Xi(p)$ is a singlet. The form of~$\Xi(p)$ can also be determined by requiring that the rank of the S~matrix $S_{12}(p_1,p_2)$ reduces (in fact, that the S~matrix has rank one) when $p_2=\bar{p}_1$, as explained in~\cite{Beisert:2015msa}.
For relativistic models $\mC$ usually coincides with the charge conjugation matrix.

We obtain crossing equations by imposing the condition
that  moving  ZF operators through the singlet does not destroy it. By using the ZF algebra, it is easy to show that
\bal
\Xi_1(p_1) \bA_2^\dagger(p_2) = \bA_2^\dagger(p_2)\,\bA_1^\dagger(p_1)S_{12}(p_1,p_2) \mC_1S_{12}^{t_1}(\bar p_1,p_2) \bA_1^{\dagger t}(\bar p_1)\,,
\eal
and therefore to get the singlet on the r.h.s. one has to impose the condition
\bal\la{creq1}
S_{12}(p_1,p_2)\, \mC_1\,S_{12}^{t_1}(\bar p_1,p_2)= \mC_1\, \cK_2(p_1,p_2)\,,
\eal
where $ \cK$ is an arbitrary matrix. If \eqref{creq1} holds then the exchange relations between the singlet and the ZF operators take the form
\bal
 \Xi_1(p_1)\, \bA_2^\dagger(p_2) = \bA_2^\dagger(p_2)\,\Xi_1(p_1)\, \cK_2(p_1,p_2)\,,\quad  \bA_1^\dagger(p_1)\,\Xi_2(p_2)=\Xi_2(p_2)\, \bA_1^\dagger(p_1)\, \cK_1^{-1}(p_2,p_1) \,.
\eal
In the usual case $ \cK$ is set to $\bI$ because  in the absence of interaction $S_{\A\B}^{\C\D}$ becomes graded identity that commutes with $\mC$. In the case of fields with non-trivial exchange relations we have\footnote{The matrix $\tilde \cR_{12}$ is obtained by first analytically  continuing $S_{12}(p_1,p_2)$ to $S_{12}(\bar p_1,p_2)$, and then taking the limit of trivial scattering. In all examples we are familiar with   $\tilde \cR_{12}=\cR_{12}$. }
\bal
S_{12}(p_1,p_2)\xrightarrow{\text{trivial scattering}} \cR_{12}(p_1,p_2)\,,\quad S_{12}(\bar p_1,p_2)\xrightarrow{\text{trivial scattering}} \tilde\cR_{12}(\bar p_1,p_2)\,.
\eal
where in the simplest case $\cR=\tilde\cR$ is a diagonal matrix. Thus, 
we get the following additional equation
\bal\la{eq:crossing-constraint}
\cR_{12}(p_1,p_2)\,\mC_1 \,\tilde \cR^{t_1}_{12}(\bar p_1,p_2)  = \mC_1\, \cK_2(p_1,p_2)\,,
\eal
which is a condition on $\mC$ and $\cR$, and it also determines $\cK$.   Note that the matrix form of $S_{12}$ depends on $\cR_{12}$, and for non-diagonal $\cR_{12}$ one has some kind of a quantum group symmetry.
Hence, in general we find the following \textit{modified crossing equation}
\begin{equation}
\label{eq:crossinggeneral}
    S_{12}(p_1,p_2)\, \mC_1\,S_{12}^{t_1}(\bar p_1,p_2)=\cR_{12}(p_1,p_2)\,\mC_1 \,\tilde \cR^{t_1}_{12}(\bar p_1,p_2)\,,
\end{equation}
subject to the constraint~\eqref{eq:crossing-constraint}; in the case $\cR_{12}$ and  $\tilde\cR_{12}$ are trivial, eq.~\eqref{eq:crossing-constraint} is trivially satisfied and~\eqref{eq:crossinggeneral} reduces to the usual crossing equation.

For a relativistic model the S-matrix and $\cR_{12}=\tilde \cR_{12}$ depend on $\theta=\theta_1-\theta_2$. The scattering in the model becomes trivial in the limit $\theta\to\pm\infty$,  and assuming $\cR_{12}$ is given by \eqref{relcR12} and using the braiding unitarity, we get
\bal
S_{12}(\theta)\xrightarrow{\theta\to\infty} R_{12}\,,\quad S_{12}(-\theta)\xrightarrow{\theta\to\infty} \left( R_{21}\right)^{-1}\,.
\eal
Thus, we get an extra compatibility condition
\bal\la{creq2}
R_{21}^{-1}\,\mC_1 \, \left(R_{21}^{-1}\right)^{t_1} =  \mC_1 \,\cK_2\,,
\eal
where $\cK$ is a constant matrix.

Let us look at the form that these non-trivial crossing equations take when expressed in terms of the ``dressing factor''~$\Sigma_i(p_1,p_2)$. Once the S-matrix is fixed by symmetries and the Yang--Baxter equation, one is left with a number of undetermined functions $\Sigma_i(p_1,p_2)$ which should be fixed by crossing, analyticity and the bound-state content of the model. If the exchange relations are trivial, the functions $\Sigma_i(p_1,p_2)$ satisfy equations of the form
\bal
\Sigma_i(p_1,p_2)\Sigma_j(\bar p_1,p_2)=f_{ij}(p_1,p_2)\,,
\eal
where $f_{ij}$ are known functions. According to the discussion above in the general case of non-trivial exchange relations the  most general self-consistent crossing equations are 
\bal
\Sigma_i(p_1,p_2)\Sigma_j(\bar p_1,p_2)=\Sigma^{(0)}_i(p_1,p_2)\Sigma^{(0)}_j(\bar p_1,p_2)f_{ij}(p_1,p_2)\,,
\eal
where $\Sigma^{(0)}_i(p_1,p_2)$ are the dressing factors in the limit of trivial scattering. In the relativistic case it is the limit $\theta_{12}\to\infty$ and the dressing factors should be chosen so that $\Sigma^{(0)}_i=\Sigma_i(\infty)$ would be constants. Then, for a relativistic model we get the most general self-consistent crossing equations in the form
\bal
\label{eq:generalised-cross-dressing}
\Sigma_i(\theta)\Sigma_j(\theta+i\pi)=\Sigma_i(\infty)\Sigma_j(\infty)f_{ij}(\theta)\,.
\eal
If the exchange relations are not known then one can look for various solutions of these crossing equations satisfying different conditions at infinity. The
minimal solution in this set would then fix the exchange relations of the model. This is what happens for the models discussed in next sections.

\subsection{Restricted Hilbert space}
\label{sec:restricted}

There are models where not all  multi-particle states are allowed, e.g.~$\Phi_{21}$ deformation of tricritical Ising model~\cite{Zamolodchikov:1990xc,Colomo:1991gw,Smirnov:1991uw}, as well as certain spin-chain models, see e.g.~\cite{Corcoran:2024ofo}. The formalism developed in this section can be easily generalised to include this type of theories. For simplicity we assume that all particles have the same dispersion relation $\w(p)$.  We can still consider  
 {\it in} and {\it out} creation and annihilation operators satisfying the  exchange relations \eqref{algcR}
 \bal\la{algcRrm}
{\bf a}_1^{\dagger} {\bf a}_2^{\dagger}&=  {\bf a}_2^{\dagger} {\bf a}_1^{\dagger}\, \cR_{12}\,,\quad {\bf a}_1{\bf a}_2=  \cR_{12}\, {\bf a}_2 {\bf a}_1 \,,\quad
{\bf a}_1 {\bf a}_2^{\dagger}=  {\bf a}_2^{\dagger} \,\cR_{21}\, {\bf a}_1 + \delta_{12}\,,
\eal
however, we do not require the R-matrix $\cR_{12}$ to be invertible on the whole space of two-particle states. The compatibility conditions for the algebra  \eqref{algcR} then lead to the following equations
 \bal
\cR_{12}^\dagger= \cR_{21} \,,\quad \cR_{12}\cR_{13}\cR_{23}=\cR_{23}\cR_{13}\cR_{12}\,.
\eal
We also require  the physical unitarity and braiding unitarity on the physical subspace of the two-particles states
\bal
\cR_{12}^\dagger\,\cR_{12}=\Pi_{12}  \,,\quad \cR_{21}\,\cR_{12}=\Pi_{12}\,,
\eal
where $\Pi_{12}$ is the projector onto the physical subspace. The restricted braiding unitarity implies 
the following constraints on $\cR_{12}$
\bal
\cR_{12}\,\Pi_{12} =\cR_{12}\,,\quad \Pi_{21}\,\cR_{12}=\cR_{12}\,.
\eal 
For definiteness we assume that the {\it in} and {\it out}  operators are chosen so that $\Pi_{12}$ is diagonal
\bal
\Pi_{12} = \sum_{\a,\B}\pi_{\A\B}\,E_\A^\A\otimes E_\B^\B\,,\quad \Pi_{21} = \sum_{\a,\B}\pi_{\B\A}\,E_\A^\A\otimes E_\B^\B
\eal
where $\pi_{\A\B}$ is equal to 1 for physical states, and 0 for unphysical ones
\bal
\pi_{\A\B} = 
\left\{
\begin{array}{ccc}
 1 & \ \text{if} \ & a_\A^{\dagger} a_\B^{\dagger} \neq 0 \\
 0 & \ \text{if}\  & a_\A^{\dagger} a_\B^{\dagger} = 0
\end{array}
\right.
\eal
Note that $\Pi_{21}=P_{12}\Pi_{12}P_{12}\neq\Pi_{12}$. The braiding unitarity condition then takes the form
 \bal\la{braidingrm}
 \sum_{\F,\G}\cR_{\F\G}^{\C\D}(p_1,p_2)\cR_{\B\A}^{\G\F}(p_2,p_1) =  \pi_{\B\A}\delta_\A^\C\delta_\B^\D\,.
 \eal
For a relativistic-invariant model we can modify \eqref{relcR12} as 
\bal\la{relcR12rm}
\cR_{12} = R_{12}\,u(\theta_{12})+ \bar R_{21}\,u(\theta_{21})\,,
\eal
where $R_{12}$ and $\bar R_{21}$ are constant  matrices 
  satisfying the YB equation, and 
 \bal
R_{12}^\dagger= R_{21} \,,\quad \bar R_{21}^\dagger= \bar R_{12}\,,\quad R_{21} \bar R_{21}= \bar R_{12}  R_{12}=  \Pi_{12}\,.
\eal
The Hamiltonian of the system in terms of the {\it in} and {\it out} creation and annihilation operators is still given by
\bal
H = \int\, dp\, \w(p)\,  a^{\dagger}_\A(p)  a^\A(p) \,,
\eal
and the eigenvectors are 
\bal
  a^{\dagger}_{\A_1}(p_1)  a^{\dagger }_{\A_2}(p_2)\cdots   a^{\dagger}_{\A_n}(p_n)|0\ket\,,
\eal
with the eigenvalues $\sum_i \w_{\A_i}(p_i)$. Note that some of the states are null.

The physical S-matrix is defined in the usual way
\bal
\bS\cdot
|p_1,p_2\ket_{\A\B}^{\rm in} =\bS_{\A\B}^{\C\D}(p_1,p_2)\,|p_1,p_2\ket_{\C\D}^{\rm in}\,,
\eal
but if the state $|p_1,p_2\ket_{\A\B}^{\rm in}$ is null (unphysical) then the S-matrix element $\bS_{\A\B}^{\C\D}$ vanish.
In the absence of interaction $\bS$ becomes the projection operator $\bS =\Pi_{12}$, and $\bS_{\A\B}^{\C\D}=\pi_{\A\B}\delta_{\A}^{\C}\delta_{\B}^{\D}$.

The ZF operators and the ZF algebra are introduced in the usual way
\bal
 \bA^{\dagger }_{1}  \bA^{\dagger}_{2} =\bA^{\dagger }_{2}    \bA^{\dagger}_{1}\, S_{12}
\,,\quad S_{12}=\cR_{12}\,\bS_{12}  \,,\quad v_1>v_2\,.
\eal
In the absence of interaction we get
\bal
S_{12}\ \xrightarrow{\text{trivial scattering}}\ \cR_{12}\Pi_{12}=\cR_{12}\,,\quad v_1>v_2\,.
\eal

\bigskip

Even though the Hamiltonian $H$ does not
 commute  with the  charges
\bal
L_\A{}^\B &= \int dp\,  a^{\dagger}_{\A}(p) a^\B(p)\,,\qquad  \A,\B=1,\ldots, K\,,
\eal 
their commutator vanishes on the two-particle physical subspace.  
Their action on two-particle states is given by the same eq.\eqref{coprod}
\bal\la{coprodrm}
   L_\F{}^\G  a^{\dagger}_{\A}(p) a^{\dagger}_{\B}(p')|0\ket &= \delta_\A^\G  a^{\dagger}_{\F}(p) a^{\dagger}_{\B}(p')|0\ket + \cR_{\A\B}^{\D \G}(p,p')\cR_{\F\D}^{\C' \C}(p',p)  a^{\dagger}_{\C}(p) a^{\dagger}_{\C'}(p')|0\ket\,,
\eal
and leads to nontrivial constraints on the form of $\cR_{\A\B}^{\C\D}$.

\bigskip

The modified crossing equations are again given by \eqref{creq1}, \eqref{creq2}
\bal\la{creq1rm}
S_{12}(p_1,p_2)\, \mC_1\,S_{12}^{t_1}(\bar p_1,p_2)= \cR_{12}(p_1,p_2)\,\mC_1 \,\tilde \cR^{t_1}_{12}(\bar p_1,p_2)\,,
\eal
subject to the condition
\bal\la{creq2rm}
\cR_{12}(p_1,p_2)\,\mC_1 \,\tilde \cR^{t_1}_{12}(\bar p_1,p_2)  = \mC_1\, \cK_2(p_1,p_2)\,,
\eal
for some $\cK_2(p_1,p_2)$,
where $ \mC$ involves the projection operator.

\section{Examples in relativistic IQFT}
\label{sec:relativistic}

In this section, we consider some examples of relativistic integrable theories with non-trivial crossing equations.
As previously explained to derive crossing we will construct certain singlet states and require these states to ``commute'' with all the other states of the ZF algebra. Here ``commute'' is in quotation marks because, a priory, we could also expect the singlet to commute, anti-commute, or have a more general braiding with the ZF operator, as we discussed in the previous section. Let us see how this works in practice on some simple examples of relativistic models.


\subsection{The \texorpdfstring{$SU(N)$}{SU(N)} chiral Gross-Neveu model}

We start considering the $SU(N)$ chiral Gross-Neveu (CGN) theory~\cite{Gross:1974jv}, whose ZF algebra is given by
\begin{equation}
\label{eq:SUN_ZF_algebra}
A_{i}^\dagger(\theta_1)A_{j}^\dagger(\theta_2) =A_{l}^\dagger(\theta_2) A_{k}^\dagger(\theta_1) S^{kl}_{ij}(\theta)\,, \qquad \theta \equiv \theta_1 - \theta_2\,,
\end{equation}
with a relativistic S-matrix fixed by symmetry considerations and unitarity to
\begin{equation}
    S_{ij}^{kl}(\theta)=\frac{\Sigma_N(\theta)}{\theta - \frac{2i \pi}{N}} \ \left(\theta \delta_i^k\delta_j^l -\frac{2i \pi}{N} \delta_i^l \delta_j^k \right)\,.
\end{equation}
The exact S-matrix of this model was originally conjectured in~\cite{Berg:1977dp,Berg:1978zn,Abdalla:1979sa,Koberle:1979ne} based on symmetry considerations combined with a large $N$ expansion. 
The physical and braiding unitarity require
\begin{equation}
    |\Sigma_N(\theta)|=1\ \text{for}\ \theta>0 \,,  \qquad \Sigma_N(-\theta)\Sigma_N(\theta)=1 \,.
\end{equation}
The model comprises both particles in the fundamental representation of $SU(N)$ and bound states, associated to completely antysymmetric representations which can be constructed as tensor products of fundamental particles. A singlet state can be obtained as a tensor product of $N$ copies of the fundamental representation as follows
\begin{equation}
\Xi(\theta_1)=\sum_{a_1, \, \dots, \, a_N=1}^N \epsilon^{a_1 \dots a_N} \prod^N_{j=1} A_{a_j}^\dagger\left(\theta_1+\frac{2 \pi i(j-1)}{N}  \right) \,.
\end{equation}
The rapidities of the constituents are fixed so that the total energy and momentum of the singlet vanish.
Scattering the singlet with any state and iterating $N$ times the commutation relation in~\eqref{eq:SUN_ZF_algebra} we obtain
\begin{equation}
\Xi(\theta_1)\, A_{k}^\dagger(\theta_2) = A_{k}^\dagger(\theta_2)\, \Xi(\theta_1)  \, \frac{\theta+ 2 i \pi \left(1 - \frac{1}{N} \right)}{\theta}\prod^N_{j=1} \Sigma_N\!\left(\theta+\frac{2 \pi i(j-1)}{N} \right) \,,
\end{equation}
where as usual we set $\theta= \theta_1 - \theta_2$. 
The requirement that the singlet must scatter as in the free theory limit $\theta \to + \infty$, which is
\begin{equation}
\begin{aligned}
\lim_{\theta \to + \infty} \Xi(\theta_1)\, A_{k}^\dagger(\theta_2) &= A_{k}^\dagger(\theta_2)\, \Xi(\theta_1)  \prod^N_{j=1}\Sigma_N\!\left(+\infty+\frac{2 \pi i (j-1)}{N} \right)\\
&
= A_{k}^\dagger(\theta_2)\, \Xi(\theta_1) \left(\Sigma_N(\infty)\right)^N\,,
\end{aligned}
\end{equation}
where in the last equation we imposed analyticity in the physical strip,
requires the following constraint to hold
\begin{equation}
\label{eq:crossing_eq_SU2_1}
\prod^N_{j=1} \Sigma_N\!\left(\theta+\frac{2 \pi i(j-1)}{N} \right) = 
\frac{\theta}{\theta+ 2 i \pi \left(1 - \frac{1}{N} \right)} \left(\Sigma_N(\infty)\right)^N\ \,.
\end{equation}
A constant phase is encoded in the value of the S-matrix at infinity through $\Sigma_N(\infty)$.  
A further requirement we should impose is that
\begin{equation}
\label{eq:collin_limit_constr}
    S_{ij}^{kl}(0)=-\delta_i^l \delta_j^k \qquad \Leftrightarrow \qquad \Sigma_N(0)=-1\,,
\end{equation}
which is necessary to have a good Bethe wavefunction.
A minimal solution for the dressing factor of the $SU(N)$ CGN model is given by
\begin{equation}
\label{eq:general_SUN_sol}
\Sigma_N(\theta)=\frac{\Gamma (+\frac{i \theta}{2 \pi}) \, \Gamma (\frac{N-1}{N}-\frac{i \theta}{2 \pi})}{\Gamma (-\frac{i \theta}{2 \pi}) \, \Gamma (\frac{N-1}{N}+\frac{i \theta}{2 \pi})} \,,
\end{equation}
with an asymptotic behaviour
\begin{equation}
\Sigma_N(\pm \infty) = e^{\mp 2 i \pi \, \frac{N-1}{2N}} \,.
\end{equation}
The non-trivial asymptotics indicates that the braiding factors for the particles in the fundamental representation of $SU(N)$ (i.e. the $s_{ij}$ terms appearing in~\eqref{diagR12}) are all equal to 
\begin{equation}
s_{ij}= \frac{N-1}{2N} \,.
\end{equation}
This yields modified crossing equations with an overall phase $e^{-i\pi (N-1)}$.

\paragraph{The case $N=2$.}
For $N=2$ we have $s_{ij}=1/4$ and if we take the limit $\theta \to + \infty$ we obtain that
\begin{equation}
\Sigma_2(\pm \infty)= \mp i \,.
\end{equation}
In other words, if we set all the braiding factors $s_{ij}=1/4$ then in the free theory limit the S-matrix becomes $-i$ and the crossing equation~\eqref{eq:generalised-cross-dressing} is
\begin{equation}
    \Sigma_2(\theta)\Sigma_2(\theta+i\pi)=-\frac{\theta}{\theta+i\pi}\,.
\end{equation}
A minimal solution to this equation was proposed in~\cite{Zamolodchikov:1992zr} to be
\begin{equation}
\label{eq:rel_Zam_Sigma_Sol}
    \Sigma_2(\theta)=\frac{\Gamma[\tfrac{1}{2}+\tfrac{\theta}{2\pi i }]}{\Gamma[\tfrac{1}{2}-\tfrac{\theta}{2\pi i }]}\frac{\Gamma[-\tfrac{\theta}{2\pi i }]}{\Gamma[+\tfrac{\theta}{2\pi i }]}\,.
\end{equation}
On the other hand, if we set $s_{ij}=0$~mod$\tfrac{1}{2}$ then we find
\begin{equation}    \Sigma_2(\theta)\Sigma_2(\theta+i\pi)=\frac{\theta}{\theta+i\pi}\,,
\end{equation}
which can in fact be solved adding an extra factor to the solution~\eqref{eq:rel_Zam_Sigma_Sol}
\begin{equation}
\label{eq:SU2_dressing_new}
\Sigma_2(\theta)=i\tanh\left(\frac{\theta}{2}-\frac{i\pi}{4}\right) \ \frac{\Gamma[\tfrac{1}{2}+\tfrac{\theta}{2\pi i }]}{\Gamma[\tfrac{1}{2}-\tfrac{\theta}{2\pi i }]}\frac{\Gamma[-\tfrac{\theta}{2\pi i }]}{\Gamma[+\tfrac{\theta}{2\pi i }]} \,.
\end{equation}
This solution is self-consistent, as 
\begin{equation}
\Sigma_2(\pm \infty)=+1 \,,\qquad
\Sigma_2(0)=-1\,, 
\end{equation}
as it should be.
One might argue that this latter solution is ``less minimal'' (because it has a zero in the physical strip at $\theta=i \pi/2$).
Still, it may be interesting to understand if it corresponds to a model of physical interest. It is amusing to note that the same solution works for $N=2,6,10,\dots$.


\subsection{\texorpdfstring{$\Phi_{21}$}{Phi-21} deformation of Tricritical Ising Model }

Let us now consider a more involved example coming from deforming the deformation of Tricritical Ising Mode (TIM) by the relevant operator $\Phi_{21}$. This model was studied in~\cite{Zamolodchikov:1990xc,Colomo:1991gw,Smirnov:1991uw}, and more recently in~\cite{Copetti:2024rqj,Copetti:2024dcz}, where the authors argued on the modification of the crossing equations suggested by the presence of non-invertible symmetries. Here we show how the result of~\cite{Colomo:1991gw,Smirnov:1991uw,Copetti:2024rqj,Copetti:2024dcz} follows from the discussion of section~\ref{sec:restricted}.

This model has two vacua: $0$ and $1$. The spectrum of the model is composed of three particles: a kink interpolating between $0$ and $1$, an anti-kink interpolating between $1$ and $0$, and a breather interpolating between $1$ and $1$.
We label these particles by $K_{01}$, $K_{10}$ and $K_{11}$. The space of two-particle states is subject to the topological constraint that the boundary vacua cannot change in the scattering. Using the same notation of~\cite{Copetti:2024rqj}, we can then write the operator ZF algebra as
\begin{equation} 
K_{dc}(\theta_1) K_{cb}(\theta_2)= \sum_{a=0,1} K_{da}(\theta_2) K_{ab}(\theta_1) S^{ab}_{dc}(\theta_{12}) \,.
\end{equation}
Note that the vacuum $c$ interpolating between the incoming states is fixed, while we are summing over all possible vacua $a$ interpolating between the outgoing states. The asymptotic vacua $d$ and $b$ cannot change since they provide the value of the topological charge.
The S-matrix element can be written in terms of the following compact expression (see e.g.~\cite{Copetti:2024rqj})
\begin{equation}
\begin{split}
&S^{ab}_{dc}(\theta)= \frac{\Sigma(\theta)}{\sinh \frac{\theta+ i \pi}{5/9}} \left( \sqrt{\frac{d_a d_c}{d_b d_d}} \sinh \frac{\theta}{5/9} \delta_{bd} + \sinh \frac{i \pi - \theta}{5/9} \delta_{ac} \right)\,,\\
&d_0=1\,, \quad d_1=\frac{1+\sqrt{5}}{2}\,.
\end{split}
\end{equation}
and the ZF algebra takes the following form
\begin{equation}
\begin{split}
K_{01}(\theta_1) K_{10}(\theta_2)&= K_{01}(\theta_2) K_{10}(\theta_1) S_{01}^{10}(\theta_{12})\,,\\
K_{01}(\theta_1) K_{11}(\theta_2)&= K_{01}(\theta_2) K_{11}(\theta_1) S_{01}^{11}(\theta_{12})\,,\\
K_{11}(\theta_1) K_{10}(\theta_2)&= K_{11}(\theta_2) K_{10}(\theta_1) S_{11}^{10}(\theta_{12})\,,\\
K_{11}(\theta_1) K_{11}(\theta_2)&= K_{11}(\theta_2) K_{11}(\theta_1) S_{11}^{11}(\theta_{12}) + K_{10}(\theta_2) K_{01}(\theta_1) S_{11}^{01}(\theta_{12})\,,\\
K_{10}(\theta_1) K_{01}(\theta_2)&= K_{10}(\theta_2) K_{01}(\theta_1) S_{10}^{01}(\theta_{12}) + K_{11}(\theta_2) K_{11}(\theta_1) S_{10}^{11}(\theta_{12})\,.
\end{split}
\end{equation}
These relations involve five (out of nine) two-particle states.
We also enlarge the algebra by adding the four conditions $K_{01}(\theta_1) K_{01}(\theta_2)=K_{10}(\theta_1) K_{10}(\theta_2)=K_{10}(\theta_1) K_{11}(\theta_2)=K_{11}(\theta_1) K_{01}(\theta_2)=0$, like in section~\ref{sec:restricted}.
The requirements of braiding unitarity and physical unitarity lead to the following constraints on the dressing factor $\Sigma(\theta)$
\begin{equation}
\Sigma(\theta) \Sigma(-\theta)= 1 \,, \qquad |\Sigma(\theta)|^2 = 1 \ \ \text{if} \ \theta > 0 \,.
\end{equation}

\paragraph{Crossing equations.}

If we want to construct a ``singlet'' in the sense of section~\ref{sec:restricted}, we must first demand that such a state has zero topological charge (we can assign a $\mathfrak{u}(1)$ charge $+1$ to the kink, $-1$ to the antikink, and~$0$ to the breather).
That is to say, such states are constructed of copies of particles and anti-particles: either  $K_{01}$ and $K_{10}$,  or $K_{11}$  and itself. 
We can therefore write a linear combination of the form
\begin{equation}
    \Xi(\theta)=c_1 K_{01}(\theta) K_{10}(\theta+i \pi)+c_2 K_{11}(\theta) K_{11}(\theta+i \pi)+ c_3 K_{10}(\theta) K_{01}(\theta+i \pi) \,,
\end{equation}
and fix the ratio of the three arbitrary coefficients~$c_i$ by requiring
\begin{equation}
    \Xi(\theta_1) \, K_{ab}(\theta_2) =K_{ab}(\theta_2) \, \Xi(\theta_1)\ f(\theta_{12})\,,
\end{equation}
for some $f(\theta)$. This yields
\begin{equation}
    \Xi(\theta)= \sqrt{d_1}K_{01}(\theta) K_{10}(\theta+i \pi)+K_{11}(\theta) K_{11}(\theta+i \pi)+ \frac{1}{\sqrt{d_1}} K_{10}(\theta) K_{01}(\theta+i \pi) \,.
\end{equation}
We expect this particular linear combination to be equivalent to demanding that  $\Xi(\theta)$ is a singlet under a suitable symmetry --- presumably the quantum group symmetry of~\cite{Smirnov:1991uw} and the  categorical symmetry of~\cite{Chang:2018iay,Copetti:2024rqj,Copetti:2024dcz}.
Explicitly, we find from the ZF relations
\begin{equation}
\Xi(\theta_1) \, K_{ab}(\theta_2)= K_{ab}(\theta_2) \, \Xi(\theta_1) \ \Sigma(\theta_{12}) \Sigma(\theta_{12} + i \pi) \ \frac{\sinh \left( \frac{i \pi- \theta_{12}}{5/9}\right)}{\sinh \left( \frac{2i \pi+\theta_{12}}{5/9}\right)}  \,.
\end{equation}
In the limit $\theta \to + \infty$ the expression above becomes
\begin{equation}
\Xi(\theta_1) \, K_{ab}(\theta_2)= K_{ab}(\theta_2) \, \Xi(\theta_1) \ \Sigma(+ \infty) \Sigma(+ \infty + i \pi) \ e^{-i \frac{2}{5} \pi}\,.
\end{equation}
The generalised crossing equation~\eqref{creq1rm} reads therefore
\begin{equation}
\Sigma(\theta) \Sigma(\theta + i \pi)   = \Sigma(+\infty) \Sigma(+ \infty + i \pi)  \ e^{-i \frac{2}{5} \pi} \frac{\sinh \left( \frac{2i \pi+\theta}{5/9}\right)}{\sinh \left( \frac{i \pi- \theta}{5/9}\right)} \,.
\end{equation}
A solution to this equation is given by
\begin{equation}
\Sigma(\theta)= - f_{-\frac{2}{5}}(\tfrac{9}{5} \theta)\; f_{\frac{3}{5}}(\tfrac{9}{5} \theta)\; F_{-\frac{1}{9}}(\theta)\; F_{\frac{2}{9}}(\theta)
\end{equation}
where we used the building block notation
\begin{equation}
f_\alpha(\theta)= \frac{\sinh \left( \frac{\theta}{2} + \frac{i \pi \alpha}{2} \right)}{\sinh \left( \frac{\theta}{2} - \frac{i \pi \alpha}{2} \right)} \,, \qquad F_{\alpha}(z) = -f_{\alpha}(z) f_\alpha (i \pi -z) \,.
\end{equation}
This solution was originally found in~\cite{Colomo:1991gw,Smirnov:1991uw} and disagrees with the  one originally proposed in~\cite{Zamolodchikov:1990xc} where different crossing equations were solved. The functions $F_{\alpha}(z)$ are CDD factors satisfying homogeneous crossing and their role is only to introduce the correct pole and fusion structure in the S-matrix.
With this solution, the S-matrix in the limit of equal rapidities becomes 
\begin{equation}
S^{ab}_{dc}(0)= - \delta_{ac} \,,
\end{equation}
and the only non-vanishing S-matrix elements in the ZF algebra are those associated to reflection processes.
This S-matrix has a non-trivial asymptotic behaviour:
\begin{equation}
S^{ab}_{dc}(\pm \infty)=  \left( \sqrt{\frac{d_a d_c}{d_b d_d}} e^{\mp \frac{3}{5} i \pi} \delta_{bd} + e^{\mp \frac{7}{5}i \pi} \delta_{ac} \right) \,.
\end{equation}

\section{Revisiting the \texorpdfstring{$AdS_3\times S^3\times T^4$}{AdS3xS3xT4} S-matrix}
\label{sec:adsxsxt}
Let us now consider the S-matrix of $AdS_3\times S^3\times T^4$. After lightcone gauge fixing, we are na\"ively left with a theory of eight bosons and eight fermions. Half of these excitations are related to fields on $AdS_3\times S^3$, and the other half to fields on~$T^4$. An analysis of the symmetries in the lightcone gauge-fixed theory suggests that the excitations have the dispersion relation
\begin{equation}
\label{eq:AdS3dispersion}
    H(p) = \sqrt{\left(m+\frac{k}{2\pi}p\right)^2+4h^2\sin^2\left(\frac{p}{2}\right)}\,,
\end{equation}
where $h$ and $k$ are coupling constants and $m$ is a $\mathfrak{u}(1)$ charge which distinguishes different representations, as we will summarise below.
The perturbative regime, whereby  it should be possible to recover the classical dynamics of the lightcone gauge-fixed model, corresponds to taking 
\begin{equation}
    T=\sqrt{\frac{k^2}{4\pi^2}+h^2}\to\infty\,,\qquad
    \kappa= \frac{k}{h}\quad\text{fixed},
\end{equation}
where $T$ is the string tension in dimensionless units. This can be done in different ways in different regimes of the model; for the purpose of understanding the worldsheet scattering, the most useful limit is probably the near-BMN one~\cite{Berenstein:2002jq}, whereby the momentum~$p$ is taken to be small so that $p T$ is fixed, $p=\mathsf{p}/T$. At leading order, this gives
\begin{equation}
\label{eq:dispersionBMN}
    H(\mathsf{p}) = \sqrt{\mathsf{p}^2+2mq\,\mathsf{p}+m^2}+\mathcal{O}(T^{-2})\,,\qquad
    q=\frac{\kappa}{\sqrt{4\pi^2+\kappa^2}}\,.
\end{equation}
This sort of limit has been first investigated in the context of $AdS_5\times S^5$ superstrings, where $k=\kappa=q=0$, and the mass takes integer positive values~\cite{Klose:2006zd}. The near-BMN dispersion, can be reproduced from an action obtained from the Green-Schwarz superstring action in the lightcone gauge. Moreover, corrections in $1/T$ have been reproduced from loop computations starting from that Lagrangian, reproducing, in particular, the  $\mathcal{O}(T^{-2})$ correction to the energy~\cite{Klose:2007rz}.

The near-BMN limit has also been studied in some detail for $AdS_3\times S^3\times T^4$ superstrings~\cite{Sundin:2013ypa, Engelund:2013fja, Bianchi:2014rfa,Roiban:2014cia,Sundin:2015uva,Sundin:2016gqe}. Here, the mass spectrum is different.
In particular, we have $m=0$ for the modes related to $T^4$, and $m=\pm1$ for the modes related to $AdS_3\times S^3$. An analysis of the poles of the S-matrix suggests that the $m=+ 1$ (respectively, $m=-1$) modes may form bound states with the same dispersion relation and mass~$m=+2, +3,\dots$ (respectively, $m=-2,-3,\dots$).%
\footnote{As hinted by the periodicity of~\eqref{eq:AdS3dispersion} under the simultaneous shift of $m\to m\pm k$  and $p\to p\mp2\pi$, it is possible to identify bond-states with $|m|>k$ with excitations with shifted momentum. The full structure of bound states is discussed in detail in~\cite{Frolov:2025uwz}.}
The crossing transformation relates representations with $m>0$ with representations with $m'=-m$, and vice versa, while the $m=0$ modes, which are massless, are related among themselves.
When expanding the GS action in the near-BMN limit, one finds the kinetic terms  of four real fermions  and bosons with $m=0$ (denoted by $\chi$ and $T$, respectively) along with those of the massive fields with $m^2=1$ (denoted by $\zeta$, $Y$, $Z$), with Lagrangian%
\footnote{Here we use $\partial_\pm=\partial_\tau\pm\partial_\sigma$. The indices $a,\dot{a}$ take values $1,2$ and are raised and lowered with the Levi-Civita tensor. The indices $i,j$ take values $1,2$. The interaction terms, which we omit, break relativistic invariance even when setting $q=0$.}
\begin{equation}
\label{eq:BMNquadratic}
\begin{aligned}
    &L = \frac{\delta^{jk}}{2}\left(\partial_- Z_j\partial_+ Z_k-Z_jZ_k\right)+q\varepsilon^{jk}Z_j\partial_\sigma Z_k
    +\frac{\delta^{jk}}{2}\left(\partial_- Y_j\partial_+ Y_k-Y_jY_k\right)+q\varepsilon^{jk}Y_j\partial_\sigma Y_k\\
    &\qquad +i\zeta_{\R,a}^*\left(\partial_-+iq\right)\zeta_{\R}^a+i\zeta_{\L,a}^*\left(\partial_+-iq\right)\zeta_{\L}^a+\sqrt{1-q^2}\left(\zeta_{\R,a}^*\zeta_{\L}^a+\zeta_{\L,a}^*\zeta_{\R}^a\right)\\
    &\qquad+\partial_- T^{a\dot{a}}\partial_+ T_{a\dot{a}} + i \chi^*_{\R,\dot{a}}\partial_-\chi_{\R}^{\dot a}+ i \chi^*_{\L,\dot{a}}\partial_+\chi_{\L}^{\dot a} +\text{interactions}\,.
\end{aligned}
\end{equation}
On one hand, the kinetic terms reproduce the expectation of the dispersion~\eqref{eq:dispersionBMN}. On the other hand, such an action is problematic, especially for the massless modes, because they are plagued by infrared divergences.%
\footnote{Unlike what happens in other integrable models, such as the Chiral Gross--Neveu model~\cite{Witten:1978qu}, we do not expect the massless field to acquire a mass. In fact, here masslessness is a result of a symmetry which should not be broken --- for the bosons, translation invariance on~$T^4$, while the fermions are related to them by supersymmetry. Moreover, the existence of massless fermion zero-modes is necessary to reproduce the expect spectrum of half-BPS multiplets of the model~\cite{Baggio:2017kza}.}
It was found that one- and two-loop perturbative corrections to the near-BMN expansion seemingly disagree with what expected from symmetry arguments. In particular, the correction to the dispersion relation~\eqref{eq:dispersionBMN} does not match with the expansion of~\eqref{eq:AdS3dispersion}, see~\cite{Sundin:2014ema}. This is even more striking because for $m=\pm1$ the expansion does match --- the mismatch appears only for~$m=0$.
This suggests that special care is needed when dealing with the massless sector of the theory.

The structure of the non-perturbative S matrix, as a function of the paramters $h,k$ and of the particles' momenta $p_1,p_2$, is largely fixed by symmetry arguments. In particular, one finds that the excitations introduced above transform in irreducible short representations~$\rho_m$ of a certain supersymmetric algebra, each containing two bosons and two fermions (in fact, the dispersion relation~\eqref{eq:AdS3dispersion} is a consequence of the shortening condition of this algebra). We hence have $\rho_{+1}, \rho_{-1}$ representations for the massive particles (as well as other representations $\rho_{\pm2}, \rho_{\pm3}, \dots$ for bound states thereof) and \textit{two short irreducible representations for gapless modes}, $\rho_0\oplus\rho_0$.
There are two additional symmetries, related to the indices $a=1,2$ and $\dot{a}=1,2$ in~\eqref{eq:BMNquadratic}. They come from $\mathfrak{so}(4)\cong \mathfrak{su}(2)_\bullet\oplus \mathfrak{su}(2)_\circ$, which emerges from the $T^4$ directions. Specifically, we associate undotted indices to~$\mathfrak{su}(2)_\bullet$ and dotted ones to $\mathfrak{su}(2)_\circ$. This $\mathfrak{so}(4)$ symmetry is not a symmetry of the spectrum --- it is broken by the boundary conditions of the massless fields --- but it appears naturally in the GS action because spinors are charged under this $SO(4)$. Because the S~matrix is insensitive to boundary conditions, it is natural to assume that it preserves such an $\mathfrak{so}(4)$ symmetry. As it can be seen from the Lagrangian, $\mathfrak{su}(2)_\bullet$ and $\mathfrak{su}(2)_\circ$ play quite different roles. While $\mathfrak{su}(2)_\bullet$ acts as an automorphism on the supersymmetry generators, $SU(2)_\circ$ commutes with them. The two massless irreducible representations $\rho_0\oplus\rho_0$ are charged under $SU(2)_\circ$, and make a doublet. This symmetry structure fixes much of the two-particle S~matrix.

Let us consider the scattering of the two massless representations among themselves. The representation theory of $\rho_0$ fixes almost entirely, up to an overall factor, the scattering  of each irreducible block in terms of a matrix $\mathbb{S}^{\text{susy}}$; because $\rho_0$ is four dimensional, the block $\mathbb{S}^{\text{susy}}$ is $16\times 16$; its explicit form was first found in~\cite{Borsato:2014hja,Lloyd:2014bsa}. Moreover, these blocks must be reshuffled in a way compatible with $\mathfrak{su}(2)_{\circ}$ symmetry, so that we have, for the scattering of massless particles,~\cite{Borsato:2014hja}
\begin{equation}
    S(p_1,p_2) = \Sigma_{\circ\circ}(p_1,p_2)\; S^{\mathfrak{su}(2)}(p_1,p_2)\otimes S^{\text{susy}}(p_1,p_2)\,,
\end{equation}
where $\Sigma(p_1,p_2)$ is a dressing factor and $S^{\mathfrak{su}(2)}(p_1,p_2)$ is fixed by requiring that it satisfies the crossing equation, so that
\begin{equation}
    S^{\mathfrak{su}(2)}(p_1,p_2)_{\A\B}^{\C\D}=
    \frac{w(p_1)-w(p_2)}{w(p_1)-w(p_2)-i}\delta_\A^\C\delta_\B^\D+\frac{-i}{w(p_1)-w(p_2)-i}\delta_\A^\D\delta_\B^\C\,,
\end{equation}
for some suitable rapidity~$w(p)$. This is the same structure as for the $\mathfrak{su}(2)$ CGN model. In this non-relativistic model, crossing corresponds to the transformation
\begin{equation}
    p\to\bar{p}=-p\,,\qquad H(p)\to H(\bar{p})=-H(-p)\,.
\end{equation}

The crossing equations for the model were derived in~\cite{Borsato:2014hja,Lloyd:2014bsa} under the assumption that, in the language of the previous sections, $\mathcal{R}_{\A\B}^{\C\D}=\delta_\A^\C\delta_\B^\D$.
It is reasonable to assume that $\mathcal{R}_{\A\B}^{\C\D}$ is diagonal within the supersymmetric representations, because the representation structure is closely related to that of massive representations --- and to that of the $AdS_5\times S^5$ model --- where a posteriori one can check that the exchange relations match what na\"ively expected from the BMN Lagrangian~\eqref{eq:BMNquadratic}.
Such a check is much harder for massless modes: first of all, it is not at all clear what the ``trivial scattering'' limit of the S~matrix should be. One guess would be the BMN limit, which however gives massless relativistic interacting bosons and fermions, which are badly behaved in the IR; moreover, the near-BMN limit of the S~matrix predicts a non-trivial collinear scattering, which cannot make sense in a perturbative theory. Other limits, like setting $k=0$ and taking $h\to0$ (which should correspond to a free limit in the dual gauge theory) also do not have an obvious physical interpretation from the point of view of the worldsheet S~matrix.
The simplest and arguably most natural way that a non-trivial exchange relation may appear is along the lines of what we saw for the $SU(2)$ CGN model, where the $SU(2)$ block obeys non-trivial exchange relations and the crossing equation is modified by a phase $e^{\mp 2i\pi s_{jk}}=-1$.
Then the crossing equations are%
\footnote{In the notation of~\cite{Frolov:2021fmj}, $\Sigma_{\circ\circ}=(\sigma^{\circ\circ})^2$ and $f_{\circ\circ}=-f^2$.}
\begin{equation}
\Sigma_{\circ\circ}(\bar{p}_1,p_2)\Sigma_{\circ\circ}(p_1,p_2)= - \frac{w(p_1)-w(p_2)}{w(p_1)-w(p_2)+i} f_{\circ\circ}(p_1,p_2)\,,
\end{equation}
where $f_{\circ\circ}(p_1,p_2)$ is the result of~\cite{Borsato:2014hja,Lloyd:2014bsa} for the crossing within the supersymmetric block (whose explicit form we omit here). Note the explicit minus sign due to the non-trivial braiding (which we have taken to be one of the $SU(2)$ CGN) and the standard $SU(2)$ crossing written in terms of $w(p)$. This equation is supplemented by the condition
\begin{equation}
    w(\bar{p})=w(p)+i\,,
\end{equation}
which comes from the off-diagonal part of the crossing equation and indicates that~$w(p)$ should have a logarithmic monodromy.
Because perturbatively it is found that~\cite{Sundin:2015uva}
\begin{equation}
    \log\frac{\mathbb{S}^{\mathfrak{su}(2)}(p_1,p_2)_{12}^{21}}{\mathbb{S}^{\mathfrak{su}(2)}(p_1,p_2)_{11}^{11}}=\log\frac{i}{w(p_1)-w(p_2)+i}=\mathcal{O}(T^{-3})\,,
\end{equation}
it appears impossible to have a non-trivial $w(p)$ and we are left with the option $w(p)=\infty$,%
\footnote{This is the only choice consistent with the factorisation of the symmetry algebra expected from the ``hexagon'' formalism~\cite{Eden:2021xhe} for the computation of correlation functions.}
which is tantamount to setting
\begin{equation}
    \mathbb{S}^{\mathfrak{su}(2)}(p_1,p_2)_{\A\B}^{\C\D}=
    \delta_\A^\C\delta_\B^\D\,,\qquad
    {S}^{\mathfrak{su}(2)}(p_1,p_2)_{\A\B}^{\C\D}=
    \delta_\A^\C\delta_\B^\D\,e^{\mp i\frac{\pi}{2}\text{sgn}(v_1-v_2)}.
\end{equation}
The new ingredients, with respect to~\cite{Borsato:2014hja,Lloyd:2014bsa}, are the non-trivial braiding (which makes it important to distinguish the ZF S matrix $S$ from the physical one~$\mathbb{S}$) and the minus sign in 
\begin{equation}
    \Sigma_{\circ\circ}(\bar{p}_1,p_2)\Sigma_{\circ\circ}(p_1,p_2)= -f_{\circ\circ}(p_1,p_2)\,.
\end{equation}
The solution of these equations will differ from the one discussed in~\cite{Frolov:2021fmj} by a function of the momenta. This function was written as $a(\gamma(p_1)-\gamma(p_2))$ in~\cite{Frolov:2021fmj}, and it is precisely of the form which we discussed for the $SU(2)$ CGN model, $a(z)=-i\tanh(\tfrac{1}{2}z-\tfrac{i}{4}\pi)$ cf.\ eq.~\eqref{eq:SU2_dressing_new}.
The new solution of crossing amounts to dropping $a(\gamma(p_1)-\gamma(p_2))$, and it is compatible with near-BMN perturbation theory for the physical S~matrix~$\mathbb{S}$.

This model also features non-trivial scattering between massless and massive ($m\neq0$) excitations. It is possible (even suggestive) to assume similar non-trivial exchange relations between massive and massless representations. This does not yield such a substantial modification of the dressing factor, but it would result in changing the massless-massive and massive-massless ZF S~matrices by $\pm i$. This would not affect the physical S~matrix, and hence the matching with perturbation theory. As for the massive-massive, also in analogy with $AdS_5\times S^5$ superstrings, we do not see a rationale to introduce non-trivial exchange relations.

\section{Conclusions}

In this paper we proposed modified crossing equations for two-dimensional integrable models with fields subject to non-trivial exchange relations. We have shown that these  equations hold for the $SU(N)$ CGN model, and the $\Phi_{21}$ deformation of TIM. The application of the equations to the $AdS_3\times S^3\times T^4$ case lets us to conjecture new crossing equation for the massless excitations. The resulting dressing factor has simpler analytic structure than the one proposed in~\cite{Frolov:2021fmj}, more akin to the structure that can be surmised from the QSC~\cite{Ekhammar:2024kzp}.
Our new proposal is self-consistent but it would be desirable to establish a first-principle derivation of the non-trivial exchange relations which we postulated, perhaps from a quantum-group symmetry (cf.~\cite{Smirnov:1990pr}) or a categorical symmetry (cf.~\cite{Copetti:2024rqj}).

Furthermore, it would be interesting to apply our logic and the generalised crossing equation~\eqref{eq:crossinggeneral} to other realizations of the ZF algebra, relativistic and non-relativistic. This may lead to the discovery of new interesting models, whose properties should then be explored by the powerful arsenal of the theory of integrable models.

\section*{Acknowledgments}

We thank the participants of the Workshop \textit{Workshop on Higher-d Integrability} in Favignana (Italy) in June 2025 for stimulating discussions, and in particular M.~de Leeuw, P.~Dorey, G.~Mussardo, N.~Primi, and R.~Tateo. We also thank F.~Ambrosino, P.~Marchetti,  and R.~Volpato for insightful discussions.
A.S.~acknowledges support from the PRIN Project n.~2022ABPBEY, from the CARIPLO Foundation under grant n.~2022-1886, and from the CARIPARO Foundation Grant under grant n.~68079.
This work has received funding from the Deutsche Forschungsgemeinschaft (DFG, German Research Foundation) -- SFB-Gesch\"aftszeichen 1624 -- Projektnummer 506632645. 
S.F.~acknowledges support from the INFN under a Foreign Visiting Fellowship.

\bibliographystyle{JHEP}
\bibliography{refs}
\end{document}